\documentclass[epj]{svjour}

\usepackage{amsmath}
\usepackage{amssymb}
\usepackage{amsfonts}
\usepackage{graphicx}
\usepackage{bm}

\begin{document}

\title{Stability and dynamical properties of material flow systems on random networks}
\author{Kartik Anand\inst{1} \and Tobias Galla\inst{2}}

\institute{Department of Mathematics, King's College London, Strand, London WC2R2LS, UK \and
School of Physics and Astronomy, The University of Manchester, Manchester M139PL, UK}

\date{Received: 21 October 2008}

\abstract{The theory of complex networks and of disordered systems is used to study the stability and dynamical properties of a simple model of material flow networks defined on random graphs. In particular we address instabilities that are characteristic of flow networks in economic, ecological and biological systems. Based on results from random matrix theory, we work out the phase diagram of such systems defined on extensively connected random graphs, and study in detail how the choice of control policies and the network structure affects stability. We also present results for more complex topologies of the underlying graph, focussing on finitely connected Erd\"os-R\'eyni graphs, Small-World Networks and Barab\'asi-Albert scale-free networks. Results indicate that variability of input-output matrix elements, and random structures of the underlying graph tend to make the system less stable, while fast price dynamics or strong responsiveness to stock accumulation promote stability.
}
\maketitle

\section{\label{sec: Introduction}Introduction}

The goals of economic policy-makers include the promotion of economic growth and minimising the effects of down-turns. Consequently, understanding the causes of business cycles or fluctuations is vital to their efforts. Research into these causes has itself followed a cyclic pattern, often peaking soon after a major economic downturn \cite{Abel:2003}.

A successful modern theory in this endeavour is the real business cycle (RBC) theory \cite{Plosser:1989}, according to which productivity
shocks (i.e., changes in oil prices, technology and management strategy) induce fluctuations in capital accumulation, consumption and
other economic indicators. Hence, the supply side of the economy is responsible for business cycles. Another major school of thought
concerning business cycles is the New-Keynesian (NK) view \cite{Farmer:1994}, where consumer and investor pessimism are responsible for business fluctuations. A significant requirement on both RBC and NK views of economic cycles is that they must be `consistent with the micro-foundations of the macro-economy' \cite{Mankiw:1992}, i.e., that the study of global quantities needs to be linked to the behaviour of the microscopic constituents of the economy under consideration.

In this regard, statistical mechanics is an invaluable toolbox, as its approach rests on deriving laws for the macroscopic behaviour of many-body systems from their microscopic rules of engagement. Research along these lines has led to a theory for complex networks \cite{Barabasi:2002,Mendes:2003,Newman:2006}. Herein, the attention has - to a certain degree - focused on the interplay between the topology of underlying interaction graphs and the robustness of systems of interacting agents. The broad scope of these studies includes ecological networks \cite{Drossel:2003}, metabolic networks \cite{Alon} as well as studies of man-made networks such as the internet \cite{Bornholdt}.

In \cite{Helbing:2004} Helbing et. al. introduce a non-linear model for material flow between sectors of an economy. In their model, each
material, or good $i$ is characterised by its inventory level, which, by virtue of a flux balance assumption, depends on the utilisation of
good $i$ by other sectors in the economy and on the rate with which this good is consumed by external agents. Consumption rates, in turn,
are influenced by the price of the goods via a non-linear demand function. The investigation of this model in \cite{Helbing:2004} is based on a linear stability analysis of the stoichiometric sectorial utilisation matrix, using empirical data. The study concludes that the model does not require exogenous shocks to explain economic fluctuations. Moreover, the arrangement of economic units in a coupled network may lead to
(undesired) global oscillations unless suitable countermeasures are taken, i.e. unless control policies are implemented which avert global
instability and fluctuations. Other models of production networks include \cite{Ponzi:2003,Helbing:2004b,Ponzi:2006,Weisbuch:2005,Weisbuch:2005b,Koenig:2007,Schweitzer:2008}.

The aim of this present paper is to broaden the scope of the findings in \cite{Helbing:2004} by considering ensembles of random input-output matrices instead of one single sample of empirical data. We focus on different ensembles of randomly assigned stoichiometric input-output matrices, and study the stability of a model whose dynamical properties are related to the ensemble of random matricies. As normally done, we map stability onto the eigenvalue spectra of the stoichiometric matrices. For networks with an extensive number of connections per node, the spectra can be found analytically using techniques from random matrix theory \cite{Mehta}. The stability of flow networks defined on finitely connected Erd\"os-R\'eyni (ER) graphs, small-word networks (SWN) and scale-free Barab\'asi-Albert (BA) networks is addressed from numerical diagonalization of the corresponding stoichiometric matrices.

The remainder of the paper is organised as follows. In
Sect. \ref{sec:Model Definition} we introduce the 
model. Sect. \ref{sec:Model sols} then outlines the steps involved in
characterising the system's eigenvalues and hence its stability. This
is followed by a detailed description of our results in Sect. \ref{sec: Results}. Finally, in Sect. \ref{sec: Conclusion} we provide concluding remarks and describe possible extensions for further work. Technical
details concerning the calculation of the density and support of
eigenvalues of Gaussian random matrices with extensive connectivity are presented in appendix \ref{apx:RMT} for completeness.

\section{\label{sec:Model Definition} Model Definitions}
\subsection{Set-up of dynamical control policies}
In this section we begin by describing the statistical model of
material flow networks, which is based on the model introduced in
\cite{Helbing:2004}. One considers a system consisting of
$i\,=\,1,\ldots,N$ units of production. For each unit
$i$, we denotes by $Q_i(t)\,\geq\,0$ its rate of operation, given in
units of delivery cycles per unit time. We also associate a unique
commodity, denoted as good $i$, with each unit. Each unit interacts with other units in the network by (i) producing/delivering goods and (ii) receiving/consuming goods from other units. The net flow, at
time $t$, of good $i$ through unit $j$ is $-a_{ij}\,Q_j(t)$, where
$a_{ij}\in\mathbb{R}$ is the difference between the amount of
good $i$ received minus the amount of good $i$ produced by unit $j$, per delivery
cycle. The sign of the $a_{ij}$ is here chosen as in
\cite{Helbing:2004}. The matrix $\mathbf{A}$ is related to the aggregate
Leontief input-output matrix \cite{Leontief:1986}. The quantity $-a_{ij}Q_j(t)$ may be positive (net production) or negative
(net consumption). We will set $a_{ii}=-1$ in the following, reflecting the assumption that unit $i$ produces one unit of good $i$ per production cycle. If we assume that a fraction, $Y_i(t)\,\geq\,0$ of
good $i$ is consumed by sinks outside the system, then the stock of good $i$
at time $t$, i.e., $S_i(t)$ is subject to the conservation law
\begin{equation}
\label{eq: Helbing Stock Update}
\frac{{\rm d}S_i(t)}{{\rm d}t}\,=\,-\sum_j a_{ij} Q_j(t)-Y_i(t)\,.
\end{equation}
We now focus on features that affect the rate of production of good $i$ at time $t$. As per \cite{Helbing:2004} the factors are two-fold: (i) if the current stock $S_i(t)$ exceeds an optimal or equilibrium level, $S_i^0\in\mathbb{R}^{+}$, the rate of production is reduced and vice versa. In operations management literature, such a strategy, which is referred to as `Constant Work-In Process' \cite{Spearman:1990}, ensures that each economic sector (or factory) maintains and mitigates its inventory and backlog; (ii) if the rate of stock accumulation is growing, ${\rm d}S_i(t)/{\rm d}t\,>\,0$, this is an independent reason to reduce $Q_i(t)$, and vice versa.

In addition to the aforementioned factors, each unit $i$ can potentially be subject to a control
strategy that ensures if the rate of production exceeds an optimal
value $Q_i^0\in\mathbb{R}^{+}$ known a-priori, then $Q_i(t)$ will
decrease, and vice versa. This strategy rests on the assumption that each production unit is subject to a budget constraint, limiting the range of production rates at which it can operate. Putting all these features together, we obtain, similarly to \cite{Helbing:2004}
\begin{eqnarray}
\label{eq: rate of production with conwip}
\frac{1}{Q_i(t)} \frac{{\rm d}Q_i(t)}{{\rm d}t}&=&\,\nu_i\left(\frac{S_i^0}{S_i(t)} - 1\right)\,+\gamma_i\left(\frac{Q_i^0}{Q_i(t)} - 1\right)\nonumber \\
&&-\frac{\mu_i}{S_{i}(t)}\frac{{\rm d}S_i(t)}{{\rm d}t}\,,
\end{eqnarray}
where $\nu_i$, $\gamma_i$ and $\mu_i$ are sensitivity parameters. Note that ${\rm d}Q_i(t)/{\rm d}t$ is here assumed to be proportional to $Q_i(t)$ (i.e. relative changes in production rates are considered), ensuring that $Q_i(t)\geq 0$, if one starts with non-negative initial conditions.

For large economies there is an additional equilibrating mechanism,
relating to the price of good $i$ at time $t$, $P_i(t)\,\geq\,0$. As
in \cite{Helbing:2004} the factors that affect the price are taken to
be identical to those affecting the production rates, and consequently
we use
\begin{equation}
\label{eq: Helbing Prices}
\frac{{\rm d}P_i(t)}{{\rm d}t}\,=\,\frac{1}{\alpha_i}\,\frac{P_i(t)}{Q_i(t)} \frac{{\rm d}Q_i(t)}{{\rm d}t}\,.
\end{equation}
The pre-factor $1/\alpha_i$ relates to the sensitivity of price change to the factors of influence. Specifically, $1/\alpha_i$ is the price-responsiveness, i.e. low values of  $\alpha_i$ imply that prices of commodities relax quickly to their equilibrium prices, while large values of $\alpha_i$ correspond to slow relaxation. In addition, it is also assumed that $P_i(t)$ affects the consumption $Y_i(t)$ via a demand function, $f_i(P_i(t))$, which is non-linear. As per standard practice, $f_i(P_i)$ is a monotonic decreasing function of $P_i$. We will write \begin{equation}
\label{eq: Helbing Consumption}
Y_i(t)\,=\,[Y_i^0\,+\,\xi_i(t)]\,f_i(P_i(t))\,,
\end{equation}
where $\xi_i(t)\in\mathbb{R}$ are Gaussian random fluctuations and $Y_i^0\in\mathbb{R}^{+}$ is the equilibrium value of external consumption of good $i$. The demand function is modeled as
\begin{equation}
\label{eq:demand function 01}
f_i(P_i(t))\,=\,{\rm max}\,\left(0,\,d_i\,-\,\widehat{d}_i\,P_i(t)\right)\,,
\end{equation}
where $d_i$ and $\widehat{d}_i$ are non-negative real numbers. The equilibrium price is $P_i^0\in\mathbb{R}^{+}$.

We will now specify choices for the network structure, i.e. the stoichiometric coefficients $a_{ij}$, and discuss their relation to the equilibrium values $\{Q_i^0,S_i^0,P_i^0\}$. In addition to these parameters, the model is defined by the variables $\{\nu_i,\mu_i,\alpha_i,\gamma_i,d_i,\widehat d_i\}$. The $\{d_i\}$ and $\{\widehat d_i\}$ determine the response of external consumption to changes of price. The remaining variables $\{\nu_i,\mu_i,\alpha_i,\gamma_i\}$ lay out the {\em control policies} of the production units, and determine their dynamical adaptive behaviour. They are hence the key control parameters an economic policy-maker would adjust so as to maximise stability, and to minimise systemic fragility and undesired fluctuating or oscillatory behaviour.

\subsection{\label{sec:netw}Structure of interaction matrices}
The freedom to choose appropriate units for $Q_i(t)$ allows us to
re-scale the $a_{ij}$ such that, the equilibrium fixed point (FP) solution of Eq.
(\ref{eq: Helbing Stock Update}) is given by
\begin{equation}
\label{eq: stock equilibirum}
Y_i^0\,=\,-\sum_j a_{ij}\,,
\end{equation}
i.e. we scale all $a_{ij}$ such that $Q_i^0=1$ for all $i$. A similar approach was taken in \cite{Helbing:2004}. We also assume that at equilibrium the external consumption of goods is homogeneous for all goods, i.e., $Y_i^0\,=\,1$. This simplification allows us to express Eq. (\ref{eq: stock equilibirum}) as
\begin{equation}
\label{eq: Constraint 1}
\sum_j a_{ij}\,=\,-1.
\end{equation}
These conditions ensure that the overall flux of goods, including a non-negative outflow $\{Y_i\}$, is balanced, i.e. that no intrinsic creation of material occurs in the system (impossibility of the Land of Cockaigne \cite{Morishima:1965,Lancaster}).  

We furthermore assume $S_i^0\,=\,1$, i.e., the desired equilibrium stock level of good $i$ corresponds to the net outflow $Y_i^0$ of good $i$ per unit time. Again, similar assumptions pertaining to $Q_i^0=Y_i^0=S_i^0$ have been made in \cite{Helbing:2004}.

While in \cite{Helbing:2004} a specific input-output matrix,
constructed from real-world data, was considered, we here focus on a
synthetic stochastic setting, in which matrix elements $a_{ij}$ are
chosen to be random variables drawn from an ensemble. They are held fixed during the course of the temporal
evolution of the $\{Q_i(t),S_i(t),P_i(t)\}$. In the language of
disordered systems theory the matrix elements are `quenched'
\cite{Mezard:1987} variables. As mentioned earlier, $a_{ij}$ represents the
efficiency with which good $i$ is utilised to produce good
$j$. Changing $a_{ij}$ is akin to structural changes in the production
mechanism; adopting new technology, for example. It is reasonable to
assume that such changes occur on a time-scale slower than that of
our dynamical degrees of freedom, hence justifying our approach to
regard the interaction matrices as quenched random variables. The paradigm of networked systems with randomly chosen interaction graphs and coupling constants will be discussed further below.

\subsubsection{Structure of matrix elements}
We are interested in the case where each unit interacts with a
fraction of the other units. This consideration may be formalised by
decomposing
\begin{equation}
\label{eq: Decompose interactions}
a_{ij}\,=\,c_{ij}\,u_{ij}\,,
\end{equation}
where $c_{ij}\,\in\,\{0,1\}$ are quenched connectivity coefficients,
determining the adjacency matrix of the flow network, and $u_{ij}$
describes the amount of good $i$ utilised by unit $j$. 

The constraint in Eq. (\ref{eq: Constraint 1}) is satisfied by constructing the $u_{ij}$ as a linear combination of random variables $J_{ij}$. First, drawing the $J_{ij}$ from some ensemble, and taking into account $a_{ii}=-1$, we set for $i\,\neq\,j$
\begin{equation}
\label{eq:explain constraint}
a_{ij}\,=\,c_{ij}\,\underbrace{\left( J_{ij}\,-\,\frac{1}{|{\cal N}_i|}\sum_{k \in {\cal N}_i} J_{ik}\right)}_{u_{ij}}\,.
\end{equation}
The set ${\cal N}_i$ in Eq. (\ref{eq:explain constraint}) denotes the elements on row $i$ such that $c_{ij}\,=\,1$. Analogous approaches have been taken in \cite{demartino-2003,thankdemartino}. This approach breaks down, however, if the elements on each row of matrix $\mathbf{J}$ are identical. By providing a large enough variance for the distribution of $J_{ij}$ we ensure that this case is avoided. Constrained random matrices have previously also been analysed in the context of glassy relaxation \cite{PhysRevB.38.11461}. In this case, however, the formulation of the row constraint induced further correlations between off-diagonal and diagonal matrix elements. Our implementation, as discussed in Appendix \ref{apx:RMT}, avoids this issue, hence simplifying further analysis.

Below, we investigate the stability properties of the model for the cases of
(i) dilute, but extensively connected and (ii) finitely connected ER random networks, (iii) networks exhibiting the small world property and (iv) scale-free networks.

\subsubsection{\label{sec:GE} Gaussian dilute ensemble}
We assume $c_{ij}\,=\,c_{ji}$, i.e., we consider the underlying network
(as defined by the adjacency matrix) to be undirected. Directionality
in the resulting material flow is modelled by allowing the utilisation
parameters $J_{ij}$ to be asymmetric, i.e by allowing for cases in
which $J_{ij}\neq J_{ji}$, as we will detail below. The connectivity coefficients are drawn
from the following distribution:
\begin{equation}
\label{eq: Distribution of connectivity}
P(c_{ij})\,=\,\left(1\,-\,\frac{c}{N}\right)\delta_{c_{ij},0}\,+\,\frac{c}{N}\delta_{c_{ij},1}\,,
\end{equation}
where $c\in\mathbb{R}^{+}$ is the average connectivity per production unit.

In what follows we consider the limit of so-called `extreme dilution'
\cite{kuhn:2006}, where $c\,\to\,\infty$ and $N\,\to\,\infty$, while the
ratio $c/N$ tends to zero, i.e. $c/N\,\to\,0$. This may be achieved by
allowing, for example $c\sim {\cal O}(\log N)$. This assumption has important implications for the
structure of the adjacency graph. Firstly, each node in the graph will
be connected to a vanishing fraction of the total number of
nodes. Secondly, the length of a typical loop is ${\cal O}(\log N)$
\cite{Derrida:1987}. Thus, taking $N\,\to\,\infty$, the probability of
finding loops of finite length tends to $0$. The environment about
each node is thus locally tree-like.

The utilisation parameters $J_{ij}$ are also taken to be
quenched. To allow for a well-defined thermodynamic limit,
$N\,\to\,\infty$ and $c\,\to\,\infty$, the mean and variance of
${J}_{ij}$ need to scale suitably with $c$. Specifically,
\begin{equation}
\label{eq:define J_ij}
{J}_{ij}\,=\,\frac{J}{\sqrt{c}}x_{ij}\,,
\end{equation}
leads to a mathematically non-trivial regime, where we choose the $x_{ij}$ to be Gaussian random variables of zero mean and a variance of order ${\cal O}(N^0)$ \cite{anand:016111}. Specifically, the $x_{ij}$ are
independent in pairs and for $i\,\neq\,j$ and $k\,\neq\,l$ and have the following moments:
\begin{equation}
\label{eq:define x_ij}
\langle x_{ij} \rangle\,=\,0\,, \qquad \langle x_{ij}\,x_{kl} \rangle\,=\,\delta_{ik}\,\delta_{jl}\,+\,\Gamma\,\delta_{il}\,\delta_{jk}\,.
\end{equation}
The parameter $\Gamma\,\in\,[-1,1]$ describes the degree of
correlations, with fully symmetric
interactions given by $\Gamma\,=\,1$. The eigenvalue spectra of such Gaussian random matrices
can be computed fully analytically, see for example
\cite{Sommers:1988} for results regarding fully connected Gaussian
ensembles. The extension to the dilute, but extensively connected case
respecting the constraint of Eq. (\ref{eq: Constraint 1}) is
straightforward, we report some steps of the corresponding calculation
in Appendix \ref{apx:RMT}. Our investigation of the stability
properties of such flow networks can hence be carried out analytically
to a large degree.

\subsubsection{\label{sec:GEFC} Gaussian finitely-connected ensemble}
We consider $c_{ij}$ to be distributed according to Eq. (\ref{eq:
Distribution of connectivity}), and continue to take
$c_{ij}\,=\,c_{ji}$. However, the average connectivity, $c$ scales as ${\cal O}(N^0)$, while we still consider the
thermodynamic limit $N\,\to\,\infty$. The statistics of the $J_{ij}$
are again those indicated in Eqs. (\ref{eq:define J_ij}) and (\ref{eq:define
x_ij}).

This scaling of $c$ and $J$ to be ${\cal O}(N^{0})$ complicates the analytical characterisation of the statistics of eigenvalues. Recent efforts \cite{Rogers:1988,Biroli:1999,Nagao:2007,PhysRevE.68.046109,1751-8121-41-29-295002,rogers-2008,Bianconi:2008,0305-4470-35-23-303} have lead to an implicit characterisation of eigenvalue densities in terms of population dynamical equations, often used in the context of the cavity method. In our analysis of finitely connected cases, we do not resort to such tools, but evaluate the corresponding eigenvalue statistics via explicit numerical diagonalization.

\subsubsection{\label{sec:GESW} Small world graphs}
Another ensemble we consider is that where the $c_{ij}$ define a Small-World Network (SWN) \cite{Watts:1998,newman:1999}. Under this paradigm, one starts from a network in which units are arranged on a one-dimensional lattice with periodic boundary conditions (i.e. a ring). Each unit is then connected to $2\ell$ ($\ell\in\mathbb{N}$) of its nearest-neighbours (i.e. $\ell$ neighbours to the right and $\ell$
neighbours to the left of the unit on the ring). In the context of an economy, `near', in a stylistic sense, models geographic or economic proximity, e.g. two production units within a country. The total number of undirected links in the system is $N\,\ell$.

Based on the algorithm proposed in \cite{Watts:1998}, starting with the first node and its pre-exiting $\ell$ nearest-neighbours links in a clock-wise direction, we re-wire each link with probability $\kappa$, i.e., the nearest-neighbour link is removed and replaced by a link to another randomly selected node. In an economic context these re-wired links may, for example, represent economic interactions of a given unit with units at long `distances', e.g. in a different country. This procedure is iterated for each node. At the end, the total number of links is still the same, while the number of re-wired, or long-ranged links is $\kappa\,N\,\ell$.

An alternative algorithm is that proposed in \cite{newman:1999}, where starting with the first node, for each of its pre-existing $2\ell$ neighbour-interactions, we add an additional link to another node with probability $\kappa'$. One then proceeds with the second node and so on. At the end of the procedure, the expected coordination number per node is $2\,\ell\,(1\,+\,\kappa')$.

Couplings strengths are given by $J_{ij}\,=\,J\,x_{ij}$, where $J\in\mathbb{R}^{+}$ and the moments of $x_{ij}$ is given by Eq. (\ref{eq:define x_ij}).

Once again, since $\ell$ and $J$ scale as ${\cal O}(N^{0})$, an analytical characterisation of  the statistics of eigenvalues of large SWN is difficult. We here limit ourselves to numerical diagonalization when addressing small world networks. We also compare results obtained for the two construction algorithms.

\subsubsection{Scale-free networks}
As a final example we study the model on a scale-free network. To this end we employ a growth process as proposed in \cite{Barabsi10151999} and construct the underlying adjacency matrix $c_{ij}$ as follows: the seed of the growth process is a network composed of two nodes, $i=1,2$, with $c_{12}=c_{21}=1$. At each time-step $t=3,\dots,N$ one further node is added to the network, and connects to {\em one} of the already existing nodes ($i=1,\ldots,t-1$) by preferential attachment, i.e. the probability of attaching to node $i\in\{1,\dots,t-1\}$ is proportional to the degree of node $i$. As shown in \cite{Barabsi10151999} this leads to a scale-free degree distribution $p(k)\sim k^{-3}$ asymptotically, i.e. in the limit of infinite network size.  In our simulations this scaling is reproduced faithfully, yielding e.g. exponents of $-2.9$ at system sizes of $N=1000$. In our analysis below we will use smaller networks of typically $N=100$ nodes for computational reasons (the stability analysis entails diagonalization of matrices of size $3N\times3N$ which can be costly if a large number of samples needs to be considered). For such sizes a scale-free degree distribution with a slightly smaller scaling exponent is found.
We once more take $J_{ij}\,=J\,x_{ij}$, where $J\in\mathbb{R}^{+}$ and the moments of $x_{ij}$ are given by Eq. (\ref{eq:define x_ij}).

\subsection{Paradigm of random network models}
The model as we use it here assumes that the interactions between units in the system constitute a random graph in which the presence or absence and the weight on each link existing link are  quenched random variables, i.e., drawn from some distribution and then kept fixed in time. Such random structures can, at best, be seen as a  minimalist approximation to real-world flow networks which are generally not random in their structure, and which can emerge from a growth or evolutionary process in which e.g. certain production units go `extinct' (bankrupt) and where new units join over time.

Nevertheless, studying quenched random structures allows for a meaningful abstraction of real-world phenomena and analytical tractability of the mathematical model. Such approaches have been used in a variety of different contexts such as neural networks \cite{Hopfield:1982,Coolen:2005}, economic activity \cite{demartino-2003,0305-4470-39-43-R01} and ecology \cite{May:1972,May:2001}, amongst others. In ecology in particular an ongoing debate on the effects of complexity on the stability or otherwise has been sparked by the study of random community models, and such model systems are under active investigation e.g. in \cite{Rozdilsky:2001,Jansen:2003}. The random ensembles of graph structures and distributions of couplings we use in our work are characterised by parameters such as the mean connectivity or variance of the randomly drawn elements in the Leontief matrix. The analysis thus allows for a specific characterisation of the effects of such parameters on the stability or otherwise of the system, and on its dynamical behaviour.  In subsequent work one can then build on this approach and add more realism by allowing the graph itself evolve in time \cite{0305-4470-37-31-002,Chowdhury:2005}.

A further drawback of the present model is the assumption that equilibrium values $Q_i^0, S_i^0, P_i^0$ are controlled externally (e.g. set to unity), and are not outcomes of the dynamics itself. Nevertheless, it was shown in \cite{0305-4470-39-43-R01} that the correlation of fixed-points values of the microscopic variables with coupling matrix elements in models with random interactions may often be ignored for the consideration of stability properties of random coupling models. Our lines of reasoning follow this approach.

\section{\label{sec:Model sols} Model Solutions}
Here we investigate the properties and solutions to the model presented in Sect. \ref{sec:Model Definition}. As in \cite{Helbing:2004} we henceforth assume homogeneous model parameters, $\widehat{\nu_i}\,=\,\widehat{\nu}$, $\widehat{\mu_i}\,=\,\widehat{\mu}$, $\alpha_i\,=\,\alpha$ and $\gamma_i\,=\,\gamma$ and use the demand function,
\begin{equation}
\label{eq:demand function}
f_i(P_i(t))\,=f(P_i(t))\,=\,{\rm max}\,\left(0,\,d\,-\,\widehat{d}\,P_i(t)\right)\,.
\end{equation}
At the FP we take $P_i^0\,=\,1$ and assume that $f_i(P_i^0)\,=\,1$.

\subsection{\label{subsec:LSA} Linear Stability Analysis}
We now investigate how the stability of the fixed-points depends on model parameters. To this end we perform a linear stability analysis, i.e. the eigenvalues of the Jacobian of the systems are used to characterise the dynamical behaviour of the system when subjected to external perturbations.

The linearization of the dynamical system about the fixed-points is given by
\begin{eqnarray}
\label{eq: lin s}
\frac{{\rm d}s_i}{{\rm d}t} &=& -\sum_{i=1}^{N} a_{ij}\,q_j(t)\,+\,\widehat{d}\,p_i(t)\,,\\
\label{eq: lin q}
\frac{{\rm d}q_i}{{\rm d}t} &=& -\left(\nu\,s_i(t)\,+\,\gamma\,q_i(t)\,+\,\mu\,\frac{{\rm d}s_i}{{\rm d}t}\right)\,,\\
\label{eq: lin p}
\frac{{\rm d}p_i}{{\rm d}t} &=& -\frac{1}{\alpha}\,\left( \nu\,s_i(t)\,+\,\gamma\,q_i(t)\,+\,\mu\,\frac{{\rm d}s_i}{{\rm d}t} \right)\,.
\end{eqnarray}

We note that the control policy $d$ has dropped out of our equations. While the system has dimension 3$N$, its rank is only 2$N$. Consequently, $N$ of its eigenvalues vanish. We denote the remaining 2$N$ eigenvalues by $\lambda_{i,\pm}$ with $i=1,\dots,N$. Labelling the eigenvalues of the $N\times N$ stoichiometric matrix $\mathbf{A}$ by ${\rm E}_i\,\in\,\mathbb{C}$, for $i\,=\,1\,,\ldots,\,N$, we obtain, similar to \cite{Helbing:2004}
\begin{equation}
\label{eq: eigenvalue for full system}
\lambda_{\pm,i}\,=\,\frac{1}{2}\,\left( -A_i\,\pm\,\sqrt{A_i^2\,-\,4\,B_i}\right)\,,
\end{equation}
where,
\begin{eqnarray}
\label{eq: A}
A_i &=& \mu\,\left(\frac{\widehat{d}}{\alpha}\,+\,\frac{\gamma}{\mu}\,-\,{\rm E}_i\right)\,,\\
\label{eq:B}
B_i &=& \nu\,\left(\frac{\widehat{d}}{\alpha}\,-\,{\rm E}_i\right)\,.
\end{eqnarray}

Due to the $N$ eigenvalues at zero, the FPs of our system can either be marginally stable or unstable. Such zero modes are not unusual in storage systems of the type we are considering here, and reflect the effects of an `integrating' behaviour of the buffers, which might cause the operating point to drift in time \cite{thankLammer}.  Following \cite{Helbing:2004} we hence characterise the stability or otherwise of the system in terms of the remaining $2N$ eigenvalues. Eqs. (\ref{eq: eigenvalue for full system}) - (\ref{eq:B}) relate the eigenvalues $\lambda_{\pm,i}$ of the full $3N\times 3N$ system to those of the $N\times N$ interaction matrix, $\mathbf{A}$. In what follows we show that the stability of the full system depends on the statistics of the ${\rm E}_i$ only through the support of its spectral density.

\subsection{\label{subsec:RMT} Density of eigenvalue for the Gaussian dilute ensemble}
We here establish the average density of eigenvalues, ${\rm E}_i$, $i\,=\,1,\,\ldots,\,N$, for an ensemble of $N\times N$ dilute real Gaussian random matrices, $\mathbf{A}$, defined via Eq. (\ref{eq:explain constraint}), where the $c_{ij}$'s and $J_{ij}$'s are drawn according to Eq. (\ref{eq: Distribution of connectivity}) and Eqs. (\ref{eq:define J_ij})-(\ref{eq:define x_ij}), respectively. The density of eigenvalues ${\rm E}_i$ is given by
\begin{equation}
\label{eq:DOS}
\rho({\rm E})\,=\,\left \langle\frac{1}{N}\,\sum_{i=1}^{N} \delta \left( {\rm E}\,-\,{\rm E}_i \right) \right\rangle\,,
\end{equation}
where the $\langle \ldots \rangle$ denotes an ensemble average. Following the lines of Sommers et al \cite{Sommers:1988}, and referring to the real and imaginary parts of ${\rm E}$ by $x$ and $y$, respectively (i.e. ${\rm E}\,=\,x\,+\,{\rm i}y$), we obtain
\begin{eqnarray}
\label{eq: DOS ellipse}
 \rho({\rm E})\,=\,
      \left\{ \begin{array}{l}
          \left(\pi\,a\,b\right)^{-1},\quad {\rm if}\, ((x\,+\,1)/a)^2\,+\,(y/b)^2 \leq J^2 \\ 
	\\	
	0,\quad {\rm otherwise}\,, \\
         \end{array} \right.
\end{eqnarray}
in the thermodynamic limit, $N\to\infty$. We have here written $a\,=\,1+\Gamma$ and $b\,=\,1\,-\,\Gamma$. The eigenvalues are hence uniformly distributed in the ellipse with major and minor axis given by $a$ and $b$, respectively. These results accurately match numerical results, and while we will not enter the details of the derivation of Eq. (\ref{eq: DOS ellipse}) in the main text, Appendix \ref{apx:RMT} contains some intermediate steps of the computation. In particular, with the scaling of couplings as chosen above, the result is independent of the connectivity parameter $c$.

\subsection{\label{subsec:Map Boundary} Mapping the eigenvalue support}
The stability or otherwise, and dynamical behaviour of the flow system is characterised by the eigenvalue, Eq. (\ref{eq: eigenvalue for full system}), of the $3N\times 3N$ system with the largest real part. We denoted this eigenvalue by $\lambda_{m}$. Working in the thermodynamic limit and considering the map $E_i\mapsto\lambda_{\pm,i}$, defined by Eq. (\ref{eq: eigenvalue for full system}), $\lambda_{m}$ is found to lie on the image of the {\em boundary} of the ellipse defined by Eq. (\ref{eq: DOS ellipse}). Thus in order to determine the long-term dynamical behaviour of the system only the image of this boundary needs to be considered.

\begin{figure}
        \centering
        \includegraphics[scale=0.3]{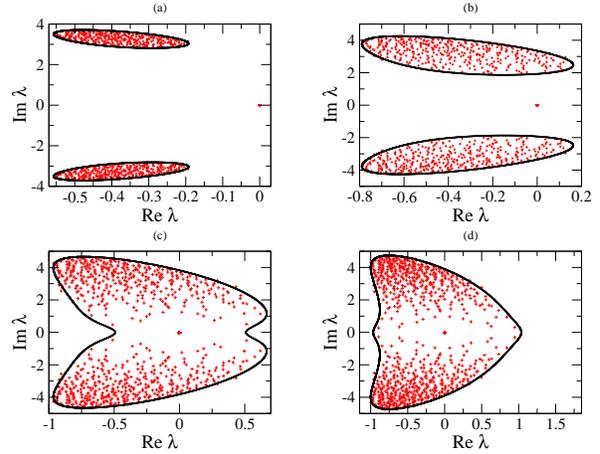}
        \caption{Eigenvalue scatter plots, wherein the crosses are results from numerical diagonalization of Eqs. (\ref{eq: lin s})-(\ref{eq: lin p}) and the black curve is from mapping eigenvalue support Eq. (\ref{eq: DOS ellipse}) via Eqs. (\ref{eq: real lambda})-(\ref{eq: img lambda}). For all plots we choose model parameters $\Gamma\,=\,0.5$, $\alpha\,=\,1$, $\mu\,=\,0.07$, $\nu\,=\,1.0$, $\gamma\,=\,0.0$ and $\widehat{d}\,=\,10.0$. In panel (a) $J\,=\,2.0$, (b) $J\,=\,5.0$, (c) $J\,=\,7.5$ and (d) $J\,=\,8.0$. System size is $N\,=\,400$.}
        \label{fig:01}
\end{figure}

Fig. \ref{fig:01} verifies the validity of this mapping and compares the analytically obtained boundary of the spectrum of the $3N\times 3N$ system against results from direct numerical diagonalization. The crosses in the figure are from numerical diagonalization of the $3N\times 3N$ system, while the solid line is from mapping of the boundary via 
\begin{eqnarray}
\label{eq: real lambda}
{\rm Re}\,\lambda_{\pm,i} &=& \frac{1}{2}\,\left( -{\rm Re}\,A_i\,\pm\,\sqrt{|D_i|}\,\cos\left( \varphi_i / 2 \right) \right)\,,\\
\label{eq: img lambda}
{\rm Im}\,\lambda_{\pm,i} &=& \frac{1}{2}\,\left( -{\rm Im}\,A_i\,\pm\,\sqrt{|D_i|}\,\sin\left( \varphi_i / 2 \right) \right)\,,
\end{eqnarray}
where $D_i\,=\,A_i^2\,-\,4\,B_i$ and $\varphi_i\,=\,\arctan \left( {\rm Im}\,D_i / {\rm Re}\,D_i \right)$.

Fig. \ref{fig:01} demonstrates that the analytical theory captures the boundary of the spectrum faithfully, and allows one to make statements regarding $\lambda_{m}$. The identification of this eigenvalue may be unique only up to complex-conjugation.

We note that the density of $\lambda_{\pm,i}$ within the predicted support is not uniform. This is due to a non-trivial Jacobian of the transformation 
\begin{equation}
{\rm p}({\rm Re}\,\lambda_{\pm},\,{\rm Im}\,\lambda_{\pm})\,=\,\rho(x,\,y)\,\left \vert \frac{\partial\,(x,\,y)}{\partial\,({\rm Re}\,\lambda_{\pm},\,{\rm Im}\,\lambda_{\pm})} \right \vert\,.
\end{equation}
However, randomly
sampling ${\rm E_i}$ according to Eq. (\ref{eq: DOS ellipse}) and
applying Eq. (\ref{eq: real lambda})-(\ref{eq: img lambda}), yields, in
the large $N$ limit, a dense scattering of $\lambda_{\pm}$ within
the mapped boundary.

The sequence of spectra shown in Fig. \ref{fig:01} reveals two
different transitions of the dynamical behaviour of the system as the
model parameter $J$, i.e. the variability of elements in the coupling
matrix, is increased.  At small $J$ (see panel (a)) one finds ${\rm Re}\,\lambda_m<0$ and ${\rm Im}\,\lambda_m\neq 0$, indicating damped
oscillations. As $J$ is increased (see e.g. panel (b)) the real part
of $\lambda_m$ becomes positive with the imaginary part still
remaining non-zero. This corresponds to growing oscillations. As $J$
is increased further $\lambda_m$ continues to display a positive real
part, but it's imaginary part vanishes, i.e., panel (d). Hence the
system is in an exponentially growing phase, where no oscillations are
to be expected. In order to verify that these eigenvalue distributions
capture the stability properties and dynamical behaviour of the system
correctly, we have integrated the linearised dynamics, Eqs. (\ref{eq:
lin s}, \ref{eq: lin q},\ref{eq: lin p}) numerically, initialising the
system close to its fixed point. Fig. \ref{fig:gdp_plot} shows the
resulting behaviour of $g=N^{-1}\sum_i (S_i(t)-S_i^0)$ as a function
of time, and results confirm the transitions predicted in by the
eigenvalue distributions shown in Fig. \ref{fig:01}. 

We discuss the phase behaviour in more detail in the next section, and focus on studying how the
different model parameters affect the stability or otherwise of the
model.

\section{\label{sec: Results} Results}
In the space of control policies, i.e., $\alpha$, $\nu$, $\mu$, $\gamma$, $\widehat{d}$, and parameters $\Gamma$ and $J$ characterising the statistics of the input-output matrix as well as in dependence on the structure of the underlying network, we ask, what policies promote stability? Due to the large number of model parameters, our investigations necessarily focus on a few specific cuts in parameter space. While this is cannot be an exhaustive enumeration of effects of all different model parameters, we find that the behaviour exhibited in these phases is rich and informative regarding the impact of policy changes.

For a given set of parameters, the system's FP is meta-stable if ${\rm Re}\,\lambda_m\,<\,0$ or is otherwise unstable. The trajectories to the FP are damped if ${\rm Im}\,\lambda_m\,=\,0$. If $\lambda_m$ has a non-zero imaginary part, then the trajectories are characterised by oscillations.

We adopt the following notation to distinguish the different phases: (i) OD: oscillatory decay (${\rm Re}\,\lambda_m\,<\,0$ and ${\rm Im}\,\lambda_m\,\neq\,0$), (ii) OG: oscillatory growth (${\rm Re}\,\lambda_m\,>\,0$ and ${\rm Im}\,\lambda_m\,\neq\,0$), (iii) ED: exponential decay (${\rm Re}\,\lambda_m\,<\,0$ and ${\rm Im}\,\lambda_m\,=\,0$) and (iv) EG: exponential growth (${\rm Re}\,\lambda_m\,>\,0$ and ${\rm Im}\,\lambda_m\,=\,0$).

Finally, for simplicity, we take $\gamma\,=\,0$. In all tested cases for $\gamma\,>\,0$ one finds that ${\rm Im}\,\lambda_m\,=\,0$. Consequently, in that case, the regions in the phase plane will be either ED or EG.

\subsection{Gaussian dilute ensemble}

\subsubsection{Preliminary observations: effects of couplings strength and symmetry of interactions}

\begin{figure}
        \centering
        \vspace{7mm}
        \includegraphics[scale=0.33]{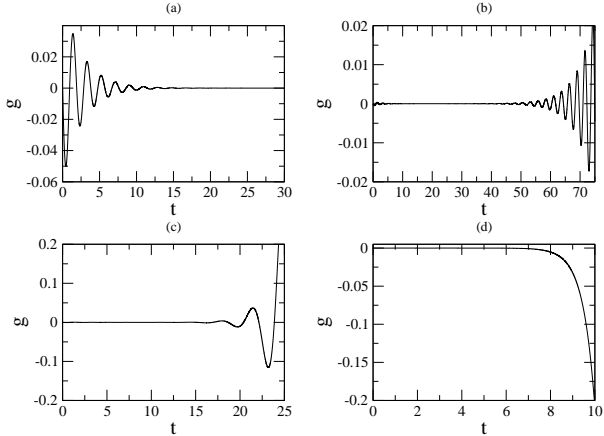}
        \caption{Deviation $g$ of the average stock accumulated in the system from the fixed point value as a function of time. Each panel corresponds to a different value of $J$: (a) $J\,=\,1.0$, (b) $J\,=\,5.0$, (c) $J\,=\,7.5$ and (d) $J\,=\,9.0$. The additional model parameters were set as $\gamma\,=\,0.0$, $\Gamma\,=\,0.5$, $\alpha\,=\,1.0$, $\mu\,=\,0.07$, $\nu\,=\,1.0$ and $\widehat{d}\,=\,10.0$. We calculated $g$ by numerically integrating Eqs. (\ref{eq: lin s})-(\ref{eq: lin p}) for $100$ units and taking the discretized time step $\Delta\,=\,0.001$.}
        \label{fig:gdp_plot}
\end{figure}

\begin{figure}
  \vspace{3em}
        \centering
        \includegraphics[scale=0.33]{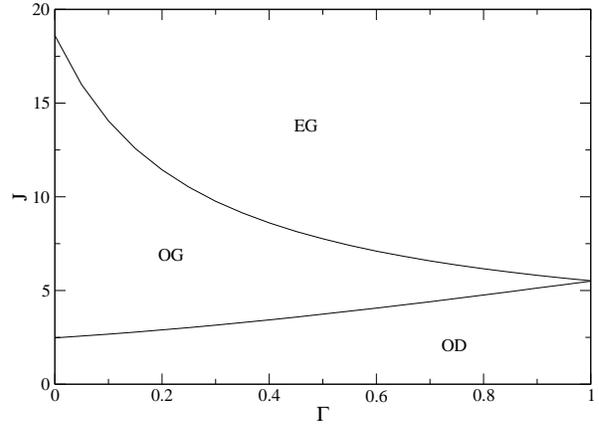}
        \caption{Gaussian dilute model: Phase boundaries in the ($\Gamma$, $J$) plane. The model parameters are: $\gamma\,=\,0.0$, $\alpha\,=\,1.0$, $\mu\,=\,0.07$, $\nu\,=\,1.0$ and $\widehat{d}\,=\,10.0$.}
        \label{fig:02}
\end{figure}

Fig. \ref{fig:02} plots the resulting phase boundaries in the ($\Gamma$,$J$) plane. Crossing the lower curve from below ${\rm Re}\,\lambda_m$ switches from negative (FP is meta-stable) to positive (FP is unstable). The upper curve separates regions with zero and non-zero ${\rm Im}\,\lambda_m$, respectively, with oscillatory behaviour found below and damped trajectories above the line. Hence, the phase space is divided into three regions, OD, OG and EG, and the transitions reported in Figs. \ref{fig:01} and \ref{fig:gdp_plot} correspond to moving along a vertical line upwards in the phase diagram, at fixed $\Gamma=0.5$.

For a given degree of symmetry between matrix elements, i.e., fixed $\Gamma$, increasing the variability between interactions, i.e., $J$, pushes the system from a stable phase with damped oscillations to an unstable phase. If the matrix elements are fully symmetric, $\Gamma\,=\,1$ then the unstable phase is always characterised by exponential growth. However, for intermediate degrees of symmetry, there exists a range of $J$ for which one observes growing oscillations. The behaviour depicted extends into the negative $\Gamma$ region. In particular, as $\Gamma$ approaches $-1$ the upper curve becomes increasingly steep and diverges for $\Gamma\,=\,-1$, where the unstable phase is always characterised by growing oscillations.

Our preliminary observations are: (i) increasing the variance of couplings makes the system more unstable and (ii) as couplings becomes more symmetric the intermediate OG phase diminishes and is absent in the fully symmetric case. In what follows we show that these findings are robust under the variation of model parameters.

\subsubsection{\label{subsec: effect of alpha}Effects of price relaxation rate}

\begin{figure}
        \centering
        \includegraphics[scale=0.33]{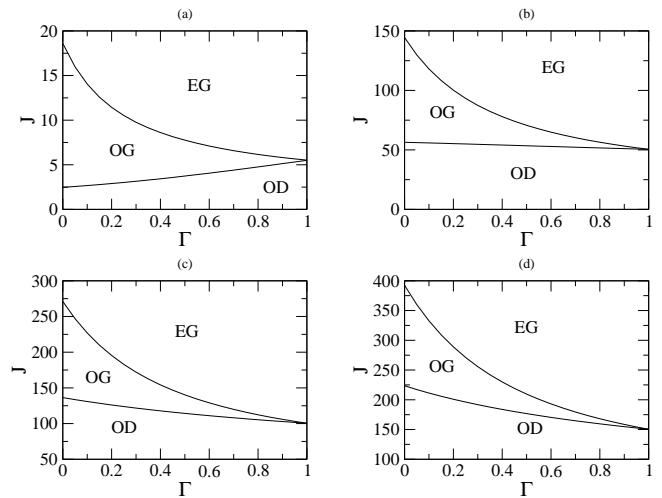}
        \caption{Gaussian dilute model: Phase boundaries in the ($\Gamma$, $J$) plane. To produce all panels, we chose fixed model parameters $\gamma\,=\,0.0$, $\mu\,=\,0.07$, $\nu\,=\,1.0$ and $\widehat{d}\,=\,10.0$. Panels differs in values of $\alpha$: (a) $\alpha\,=\,1.0$, (b) $\alpha\,=\,0.1$, (c) $\alpha\,=\,0.05$ and (d) $\alpha\,=\,0.03$.}
        \label{fig:03}
\end{figure}

Fig. \ref{fig:03} plots phase boundaries in the ($\Gamma$, $J$) plane, for different values of the price-responsiveness policy $1/\alpha$. We find that the qualitative structure of the phase diagram in Fig. \ref{fig:02} remains unchanged as $\alpha$ is varied, i.e., for different price relaxation regimes.

For a given $\Gamma$, the critical value $J^C$, separating the OD and OG phases, decreases as $1/\alpha$ is lowered. Moreover, the $J^C$ curve, as a function of $\Gamma$, switches from a monotonic increasing to a decreasing function. We conclude that quickly evolving prices, i.e., large $1/\alpha$ promotes stability.

\begin{figure}
        \centering
        \includegraphics[scale=0.33]{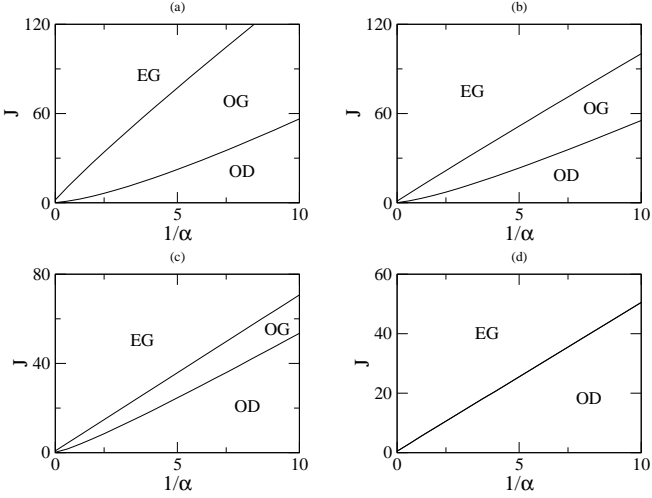}
        \caption{Gaussian dilute model: Phase boundaries in the ($1/\alpha$, $J$) plane. To produce all panels, we chose fixed model parameters $\gamma\,=\,0.0$, $\mu\,=\,0.07$, $\nu\,=\,1.0$ and $\widehat{d}\,=\,10.0$. Panels differs in values of $\Gamma$: (a) $\Gamma\,=\,0.0$, (b) $\Gamma\,=\,0.2$, (c) $\Gamma\,=\,0.5$ and (d) $\Gamma\,=\,1.0$.}
        \label{fig:04}
\end{figure}

Fig. \ref{fig:04} illustrates this further by plotting phase boundaries in the ($1/\alpha$, $J$) plane for different values of $\Gamma$.  The phase space is once again divided into the three distinct regions. The $J^C$ curve is an increasing function of $1/\alpha$. As we increase $\Gamma$ the OG regions diminish, until for $\Gamma\,=\,1$ the two phase boundaries coincide exactly with each other, excluding the OG region from the parameter space. We observe a direct transition from OD to EG. As previously mentioned, increasing $J$ moves the system from a stable to an unstable phase.

The results discussed so far are derived from the analytically known spectra of Gaussian random matrices and large connectivity in the limit of infinite system size. Only in this limit are analytical results available. In order to assess the behaviour of {\em finite} systems we consider the probability that a finite system finds itself in either of the four phases as a function of the model parameters. These probabilities were computed from a numerical diagonalization and a subsequent identification of the largest eigenvalue.

\begin{figure}
        \centering
        \vspace{3mm}
        \includegraphics[scale=0.33]{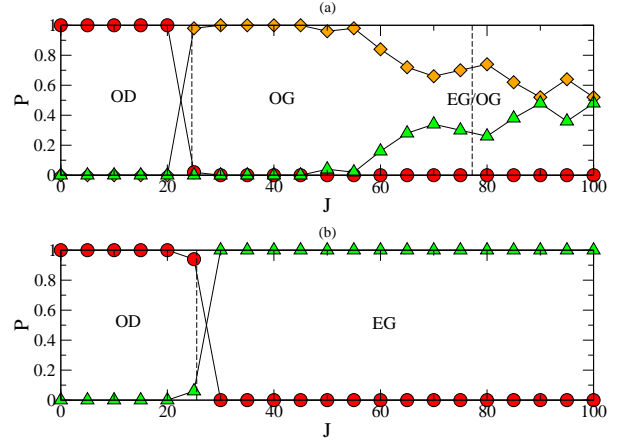}
        \caption{Gaussian dilute model: probability of finding the system in a given phase versus variability $J$ of interaction coefficients. The different symbols give the probabilities that the system finds itself in a given phase: OD (circles), OG (diamonds), EG (triangles). The remaining model parameters were $\gamma\,=\,0.0$, $\alpha\,=0.2$, $\mu\,=\,0.07$, $\nu\,=\,1.0$ and $\widehat{d}\,=\,10.0$. Panel (a) is for $\Gamma\,=\,0.0$, while panel (b) shows the case $\Gamma\,=\,1.0$. Vertical lines show phase boundaries obtained from the theory in the limit of infinite system size.  The probabilities were computed from direct diagonalization of $100$ interaction matrices at $N\,=\,100$ and mapping eigenvalues via Eqs. (\ref{eq: real lambda})-(\ref{eq: img lambda}).}
        \label{fig:05}
\end{figure}

Fig. \ref{fig:05} reports the relative frequency with which each phases occurs (phases not reported in the figure are not observed in the simulation). The behaviour of the finite system follows that predicted by the theory to a good accuracy. Discrepancies occur in a parameter regime where the theory predicts exponential growth. Here, the analytically predicted largest eigenvalue is real and positive. In finite systems however, largest eigenvalues with a non-zero imaginary part may be found due to finite-size fluctuations. We observe a similar discrepancy in the lower-right panel of Fig. \ref{fig:01}, where the theory predicts a real-valued largest eigenvalue, but explicit diagonalization at finite sizes delivers eigenvalues $\lambda_m$  with non-vanishing imaginary values. We attribute the smearing-out of the OG to EG transition and co-existence of both phases at large $J$ in the upper panel of  Fig. \ref{fig:05} to this finite-size effect.

\subsubsection{\label{subsec: effect of mu}Sensitivity of production rate to stock accumulation}

\begin{figure}
        \centering
        \includegraphics[scale=0.33]{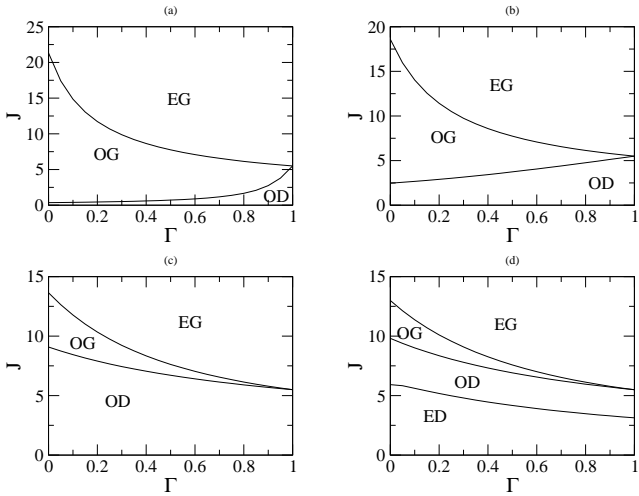}
        \caption{Gaussian dilute model: effects of control policy $\mu$: Phase boundaries in the ($\Gamma$, $J$) plane. To produce all panels, we chose fixed model parameters $\gamma\,=\,0.0$, $\alpha\,=\,1.0$, $\nu\,=\,1.0$ and $\widehat{d}\,=\,10.0$. Panels differ in the value chosen for $\mu$: (a) $\mu\,=\,0.01$, (b) $\mu\,=\,0.07$, (c) $\mu\,=\,0.5$ and (d) $\mu\,=\,0.7$.}
        \label{fig:06}
\end{figure}

The parameter $\mu$ is a measure for the sensitivity of a unit's rate of production $Q_i$ to its rate of stock accumulation $dS_i/dt$. In order to characterise the effects of this control policy, we compute in Fig. \ref{fig:06} the phase diagram in the ($\Gamma,J$) plane for different values of $\mu$. For any degree of asymmetry in the couplings, we find that increasing the sensitivity of the production rate to stock accumulation enhances stability - in panels (a)-(c) the area covered by the OD phase increases as we increase $\mu$. For $\mu\,\geq\,0.7$ we begin to observe the ED phase as well. As the sensitivity is increased further the lines separating ED-OD and OG-EG transitions converge until the system experiences a direct transition from the ED to EG phase.

Finally, we further support our findings on the roles played by $\alpha$ and $\mu$ in promoting stability, by investigating the special case of fully symmetric matrix elements, $\Gamma\,=\,1$. Here all ${\rm E}_i$ are real and in this case one can show that $\lambda_m$ is real and negative whenever the following two conditions are satisfied simultaneously:
\begin{equation}
\label{eq: alpha_c}
\alpha\,<\,\frac{\widehat{d}}{2\,J\,-\,1}\,,
\end{equation}
and
\begin{equation}
\label{eq: condition for damped behavior}
\frac{\nu}{\mu^2}\,<\,\frac{(\widehat{d}/\alpha\,\,-\,x_m)}{4}.
\end{equation}
Here $x_m$ is the real part of the eigenvalue $E$ of the Leontief matrix, giving rise to the relevant eigenvalue $\lambda_m$ determining the stability of the system. A condition similar to the latter is also found in \cite{Helbing:2004}. 
For further illustration, Fig. \ref{fig:07} shows the resulting phase diagram in the ($1/\alpha$, $\nu$/$\mu^2$) plane. The vertical line is where $\alpha_c$ is equal to the right hand side of Eq. ({\ref{eq: alpha_c}), separating the unstable EG phase at large $\alpha$ from the stable ones at smaller values of $\alpha$. If prices evolve sufficiently fast ($1/\alpha>1/\alpha_c$), increasing $\mu$ stabilises the system further (at fixed $\nu$) by supressing oscillations, as seen in Fig. \ref{fig:07}. In \cite{Helbing:2004} a related phase diagram is shown, based on stochastic modifications of an input-output matrix drawn from real world data. 

Evaluating Eq. (\ref{eq: condition for damped behavior}), which defines the phase lines separating the OD from the ED phase,  requires the knowledge of $x_m$ as a function of the model parameters. This relation may in general be somewhat intricate, at $\Gamma=1$ however one has $-2J-1\leq x_m\leq 2J-1$. The two dashed lines in the figure correspond to Eq. (\ref{eq: condition for damped behavior}) evaluated for values of $x_m$ at the limits of this interval, hence limiting the location of the phase transition separating the ED from the OD phase. The actual transition point is found to lie between these boundaries (solid line).

\begin{figure}
        \centering
        \includegraphics[scale=0.3]{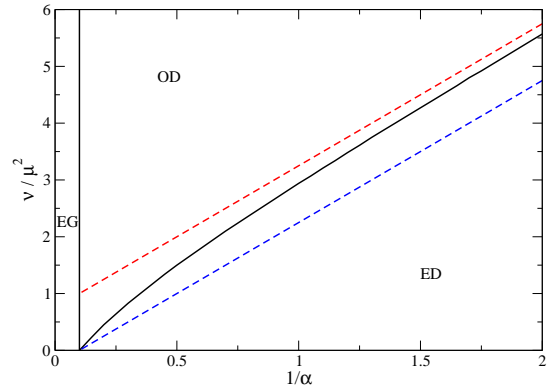}
        \caption{Dilute, but extensively connected random graph: Phase boundaries in the ($1/\alpha$, $\nu / \mu^2$) plane, for the case $\Gamma\,=\,1.0$. Additional model parameters were set as $J\,=\,1.0$ and $\widehat{d}\,=\,10.0$.}
        \label{fig:07}
\end{figure}

\subsection{\label{subsec:results_GFCE} Gaussian finitely-connected ensemble}

While the results presented for the Gaussian ensemble with a large connectivity are rich, the assumption that production networks display such topology is admittedly a very stylized one, chosen because of the available analytical results of the spectra of the corresponding couplings matrices. More realistic choices might correspond to networks in which each node is connected, on average, to a {\em finite} number of other nodes, $c$. In order to study such models we next switch our attention to finitely connected ER graphs as introduced in Sec \ref{sec:GEFC}.

Results for the density of eigenvalues for such couplings matrices are so far only available for fully symmetric couplings and have been expressed using approximation schemes, such as the single defect approximation (SDA) \cite{0305-4470-32-24-101} and effective medium approximation (EMA) \cite{PhysRevE.68.046109}. Other approaches, motivated by the statistical mechanics analysis of spin-glass type systems express the density via cavity equations \cite{rogers-2008} and as the solution to population-dynamics equations \cite{1751-8121-41-29-295002,Bianconi:2008}.

For asymmetric random matrices, however, similar results are not yet available. It has nevertheless been argued \cite{Khorunzhiy:2006} that for any finitely connected ER random graph the tails of the spectra is characterised by Lifshitz tails; hence the entire complex plane serves as the support. While most eigenvalues are found concentrated around an origin, there are outlier eigenvalues. For finite sized systems these outliers may have a significant impact on the dynamical behaviour of the system

\begin{figure}
        \centering
        \vspace{3mm}
        \includegraphics[scale=0.3]{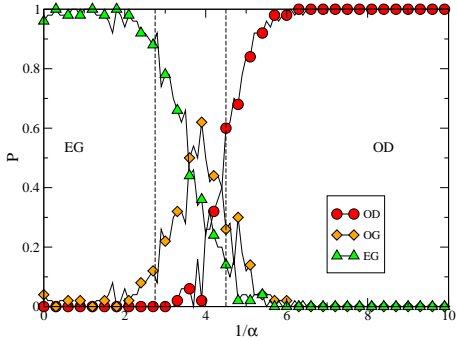}
        \caption{Gaussian finitely connected model: as a function of $1/\alpha$ we plot the probability that a particular configuration is achieved. The different symbols denote the different phases: OD (circles), OG (diamonds) and EG (triangles). For intermediate $1/\alpha$ the system is in the OG phase, but transits into the OD phase on increasing $1/\alpha$ further. The average connectivity was $c\,=\,1.0$. Additional model parameters taken were: $\Gamma\,=\,0.5$, $J\,=\,20.0$, $\mu\,=\,0.07$, $\nu\,=\,1.0$ and $\widehat{d}\,=\,10.0$. For each $\alpha$, the probabilities were evaluated from numerical diagonalization of $50$ matrices with $N\,=\,100$. }
        \label{fig:08}
\end{figure}

Fig. \ref{fig:08} plots for finite systems at an average connectivity of $c\,=\,1.0$ the probability that the system finds itself in a particular configuration, as a function of $1/\alpha$. For small $1/\alpha$, the system is unstable and in the EG phase. However, as we increase $1/\alpha$ and allow for faster price relaxation dynamics, the system first enters the unstable OG phase and later makes a transition to the stable OD phase.  When tested against $c\,=\,4$ and $c\,=\,8$ we found that the location of the transition points vary by only $1/\alpha\,=\,\pm\,0.3$. In particular, this point is in good agreement with that for the fully connected system as given in Fig. \ref{fig:04} and depicted in Fig. \ref{fig:08} by the vertical dashed lines. This is a result of our scaling of the matrix interaction term $a_{ij}$ as $1\,/\,\sqrt{c}$.

If we remove this particular scaling of the couplings the behaviour of the system becomes dependent on the mean connectivity of the underlying network,  see Fig. \ref{fig:09}. As $c$ increases, the system tends to become more unstable. In the case of full asymmetry a transition between an oscillatory decaying phase and an oscillatory unstable phase is found. If interactions are fully symmetric, then a we increase $c$ we system switches from the OD phase to a state dominated by the EG phase with high probability.

\begin{figure}
        \centering
        \vspace{5mm}
        \includegraphics[scale=0.3]{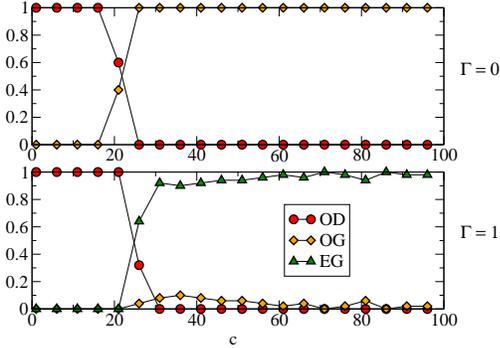}
        \caption{Gaussian model on random graphs: probability the system finds itself in a particular phase as a function of the average connectivity $c$. The different symbols give the different phases: OD (circles), OG (diamonds) and EG (triangles). Panel (a) is for $\Gamma\,=\,0.0$ while panel (b) show the case $\Gamma\,=\,1.0$. Additional model parameters taken were: $J\,=\,5.0$, $\alpha\,=\,0.2$, $\mu\,=\,0.07$, $\nu\,=\,1.0$ and $\widehat{d}\,=\,10.0$. The probabilities were computed from direct diagonalization of $50$ matrices with $N\,=\,100$. }
        \label{fig:09}
\end{figure}

\subsection{\label{subsec:results_GFSW} Small world networks}

In considering the SWNs ensemble, one introduces two new parameters to the system: (i) number of nearest-neighbours links, $\ell$ in the initial network, and (ii) the amount of long-range links, parametrised by model parameters $\kappa$ and $\kappa'$, as explained in Sec. \ref{sec:GESW}.

Under the first variant for the algorithm, \cite{Watts:1998}, each nearest-neighbour link is re-wired with probability $\kappa$. In Fig. \ref{fig:10a} we show that as $\kappa$ tends to 1, i.e., the network moves from a structured state to a random one, the system becomes more unstable. Initially the system finds itself in the OD phase, but as $\kappa$ increases we observe co-existence with the OG phase.

\begin{figure}
        \centering
        \vspace{5mm}
        \includegraphics[scale=0.3]{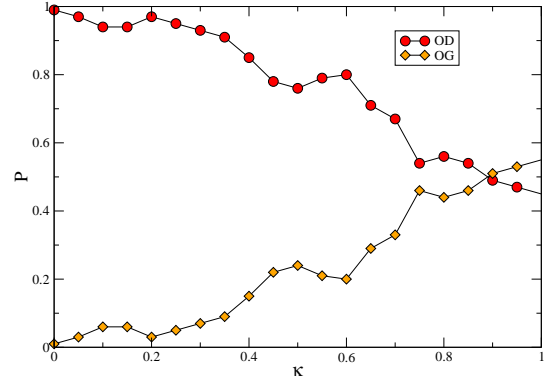}
        \caption{Small World Network: as a function of the re-wiring rate $\kappa$ we plot the probability that the system finds itself in a particular phase. The different symbols give the different phases: OD (circles) and OG (diamonds). The nearest-neighbour links $\ell\,=\,3$. Additional model parameters taken were: $J\,=\,1.0$, $\mu\,=\,0.07$, $\nu\,=\,1.0$, $\Gamma\,=\,0.0$, $\alpha\,=\,1.0$ and $\widehat{d}\,=\,10.0$. For each $\kappa$, the probabilities were evaluated from numerical diagonalization of $50$ matrices with $N\,=\,100$.}
        \label{fig:10a}
\end{figure}

For the second variant to construct the network, \cite{newman:1999}, Fig. \ref{fig:10b} plots the probability a stable state, i.e., $P_{stable}\,=\,P_{OD}\,+\,P_{ED}$ is achieved as a function of the rate $\kappa'$ with which additional links are introduced into the nearest-neighbour network. As $\kappa'$ increases, the system becomes more unstable. Similarly, stability is reduced as the number of nearest-neighbours, $\ell$, is increased. Thus, in accordance with results in Sec \ref{subsec:results_GFCE}, it is mostly the total number of links in the system that (all other parameters remaining the same) controls the stability of the network.

\begin{figure}
  \vspace{3mm}
        \centering
        \includegraphics[scale=0.3]{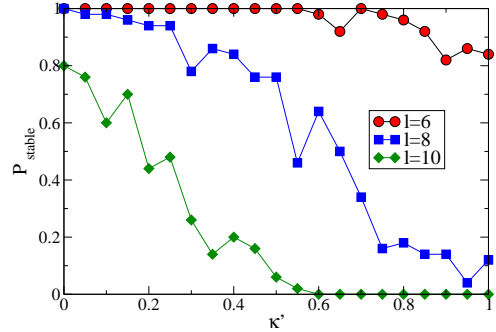}
        \caption{Small World Network: as a function of the rate $\kappa'$ with which links are added to an initially local network we plot the probability that a stable configuration is achieved. The stable state is characterised solely by the OD phase and the system makes a transition into the unstable OG phase. Additional model parameters taken were: $J\,=\,1.0$, $\mu\,=\,0.07$, $\nu\,=\,1.0$, $\Gamma\,=\,0.0$, $\alpha\,=\,1.0$ and $\widehat{d}\,=\,10.0$. For each $\kappa'$, the probabilities were evaluated from numerical diagonalization of $50$ matrices with $N\,=\,100$.}
        \label{fig:10b}
\end{figure}

\subsection{Scale-free networks}
Results for scale-free networks are shown in Fig. \ref{fig:ba} for a specific choice of parameters. We show the probability $P_{stable}$ as a function of the coupling strengths $J$ for both BA like networks and ER graphs. In both cases the model is set up such that the average degree of nodes is $<k>\approx 2$. As in previous cases, the system size is $N=100$, and results in a scaling exponent of approximately $2.4$ in the BA-case. The remaining parameters are as indicated in the figure caption. 

The upper panel reveals that for both types of the underlying adjacency matrix, the system exhibits a stable phase at low variability of elements of the Leontief-matrix, i.e. at low values of $J$ virtually all randomly drawn instances are found to be stable. Note that randomness here refers to both, the underlying graph and the coupling strengths along the links. Both systems exhibit a crossover to an unstable phase as matrix elements become more diverse. Crucially however the probability of finding a stable instance of the flow network is consistently higher in the case of BA-like graphs as compared to ER-networks. Near approximately $J\approx 2$ this effect can be significant, raising the probability of being stable from about $30$ per cent (ER) to $60$ per cent (BA). This observation is further illustrated in the lower panel of Fig. \ref{fig:ba}, where we depict the real part of the relevant eigenvalue $\lambda_m$. In the stable regime the relaxation time, or {\em resilience} of the network against perturbations is given by $\tau=1/|Re\lambda_m|$, and since $|Re\lambda_m|$ is consistently higher (in the stable phase) in the BA case as compared to the ER case we conclude that the BA network appears to be more resilient against external fluctuations than the ER graph. Similar statements can be made in the unstable phase, where $Re\lambda_m>0$ for both types of networks, but where this real-part is consistently larger in the ER case compared to the BA network. Hence perturbations to flow-networks defined on ER graphs show a much larger growth rate as compared to BA graphs. We would here like to stress that the observations presented in Fig. \ref{fig:ba} are only for one specific combination of model parameters, and that the above statements are hence valid only pending a systematic investigation of other circumstances. Still, the chose example illustrates that the structure of the underlying adjacency matrix can be relevant and that scale-free degree distributions may potentially promote stability in the context of the present model.
\begin{figure}
        \centering
        \includegraphics[scale=0.45]{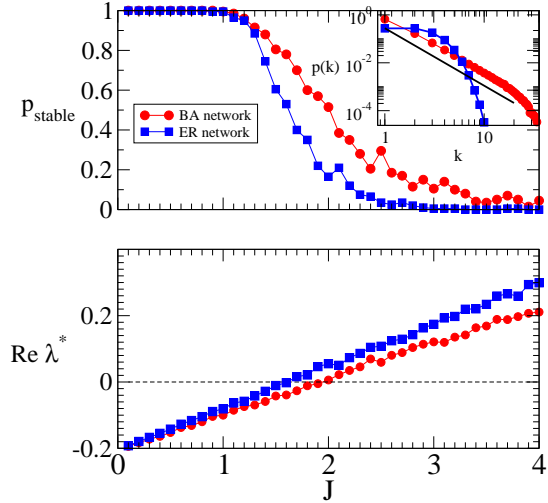}
        \caption{Behaviour of the model defined on a BA network as compared to ER graphs. Top panel: probability of finding a stable instance, i.e., $P_{\mbox{stable}}$ as a function of variability of matrix elements $J$. The lower panel depicts the expected value of the real part of the relevant eigenvalue $\lambda_m$. Inset in the upper panel shows the degree distributions of the studied adjacency matrices (solid line is $p(k)\sim k^{-2.4}$). Model parameters are $N=100,\Gamma=0, \alpha=0.2, \mu=0.01, \nu=1, \widehat d=8$. Each data point is obtained from sampling $200$ realizations of the flow-network, and subsequent direct diagonalisation. Mean degree is $c=\langle k\rangle=2$ for both considered types of networks.}
        \label{fig:ba}
\end{figure}

\section{\label{sec: Conclusion} Conclusion}
We have studied the stability and dynamical properties of a material flow system defined on a variety of random and complex network structures. Results from random matrix theory have been used to address models in the extremely dilute limit with Gaussian couplings, provided coupling strengths are scaled appropriately with the mean connectivity \footnote{However, it should be noted that the assumption needed to carry through the RMT as described in this paper is $c\,\gg\,1$ rather than sparseness. Hence, the analysis applies to networks with non-sparse connectivity with $c\,=\,{\cal O}(N)$ as well.}. Our theoretical findings are here in very good agreement with results obtained from direct numerical diagonalization, and indicate complex phase diagrams in dependence on the network structure, symmetry or otherwise of input and output matrices and most crucially on parameters characterising the external control policies applied to the network.

The analysis of Gaussian dilute networks reveals the following key findings: (i) Increasing the variance between matrix couple elements makes the system unstable, (ii) quickly evolving prices are conducive to a stable environment, (iii) greater sensitivity of the rate of production to stock accumulation can also yield greater stability by suppressing oscillations. The role of the coupling symmetry is more intricate: if prices adjust slowly or the rate of production is weakly sensitive to stock accumulation, then increasing the symmetry between the underlying network couplings makes the system more stable. For quickly evolving prices or high sensitivity, however, increasing the symmetry results in a more unstable system.

An investigation of finitely connected networks allows us to study the question of how the mean degree of nodes in the flow network affects its stability. At fixed connectivity we find that symmetry in interactions and a high degree of variability in interaction strengths reduces the extent of the stable region in parameter space, as in the fully connected model. Increased connectivity adds to the variance of coupling strengths, and hence again promotes instability.

Studying flow dynamics on SWNs reveals that an increased the re-wiring probability makes the system more unstable if prices evolve slowly. Thus, networks with a regular structure, as opposed to random networks, promote stability. Similarly, increasing the number of nearest-neighbour links has a marked impact of making the system more unstable for slow price dynamics. Finally, we have interaction graphs of BA type, and our results indicate that the underlying scale-free structures may promote stability.

The route taken in this paper was to study flow systems defined on random networks. Randomness here refers to both stochasticity in the strengths of interactions (entries in the Leontief matrix), but also to the presence or absence of links between individual units of production. The latter type of stochasticity is a common tool in the theory of complex networks, while the former has been applied in a variety of contexts, e.g. in ecology \cite{May:1972,May:2001}, linear economies \cite{andreamatteo}, evolutionary game theory \cite{Diederich_Opper_1989,Opper_Diederich_1992}. Real-world production networks are of course not random, neither in their structure nor in the magnitude of inter-unit dependencies. Still, choosing ensembles of random networks allows one to unearth general principles that determine the stability or otherwise of such models, e.g. our study consistently seems to indicate that an increased variability of elements in the Leontief input-output matrix generally induces instability. This leads to the obvious task of identifying analogous measures of variability in real-world production networks, and to verify whether or not such a correlation between complexity and stability can be confirmed.

It is also legitimate to ask what values the various model parameters would take in real-world production networks? To sensibly answer this question one must analyse real-world data, which is beyond the scope of the present paper. Statistics for $J$ and $\Gamma$ can be obtained by analysing real-world data such as used in \cite{Helbing:2004}. Positive values of $\Gamma$ indicate a tendency towards two-cycles in the production network (e.g. A uses B and B uses A). Such direct cycles are presumably unlikely and we expect small values, $\Gamma\,\approx\,0$ more realistic. Control policies such as the price responsiveness are dynamic quantities. This makes their direct measurement difficult, as the temporal behaviour of real systems would have to be probed. While calibrating the model is a necessary future step, the focus of the present paper is on the statistical mechanics analysis and phase behaviour of the model. The contribution of the present paper is a comprehensive analysis, outlining the complex interplay between relevant parameters, against which real-world scenarios may be placed.

Further directions of future research and modelling attempts include the intricate dynamics of evolving production networks, in which the underlying graph is a function of time itself, leading to a system in which discrete degrees of freedom (absence or presence of links) interacts with continuous ones (e.g. production rates defined on the nodes of the network). Such systems are known as hybrid complex systems \cite{Antsaklis:2000}. Depending on the separation of time scales, different types of dynamical behaviour might then to be expected, with the freedom of removing or adding links potentially helping to stabilise the model.

\section*{Acknowledgements}
This work was initiated at the Abdus Salam International Centre for Theoretical Physics, Trieste, Italy, which the authors would like to thank for hospitality. Fruitful discussions with Stefan L\"ammer, Reimer K\"uhn and Andrea De Martino are gratefully acknowledged. TG is an RCUK Fellow (RCUK reference EP/E500048/1).

\appendix
\section{\label{apx:RMT} Random Matrix Theory}
Following lines of reasoning provided in \cite{Sommers:1988} and employing an electrostatic analogy, the starting point to evaluate the spectra of eigenvalues is the Green's function,
\begin{equation}
\label{eq: greens function}
G({\rm E})\,=\,\frac{1}{N}\sum_{i}\frac{1}{{\rm E}_i\,-\,{\rm E}}\,,
\end{equation}
where ${\rm E}_i$ denotes the eigenvalues of matrix $\boldsymbol{A}$. The real and imaginary parts of Eq. (\ref{eq: greens function}) relate to an electric field, with charges at points ${\rm E}_i$. We can define a potential,
\begin{equation}
\label{eq: potential}
\phi({\rm E},{\rm E}^*)\,=\,-1/N\,\log\,\det \{(\boldsymbol{A}^T\,-\,{\rm E}^{*})\,(\boldsymbol{A}\,-\,{\rm E})\}\,,
\end{equation}
where ${\rm E}^*$ is the complex-conjugate of eigenvalue ${\rm E}$. Eq. (\ref{eq: potential}) satisfies, for ${\rm E}\,\neq\,{\rm E}_i$, $\partial \phi / \partial {\rm E}\,=\,G({\rm E})$. The average density of eigenvalues $\rho$ was shown to be related to $\phi$ averaged over an ensemble of matrices $\boldsymbol{A}$, i.e., the disorder average, $\langle\,\phi\,\rangle$, via Poisson's equation,
\begin{equation}
\label{eq: Poission eq}
\rho\,=\,-\frac{1}{4\,\pi}\bigtriangledown\langle\,\phi\,\rangle\,,
\end{equation}
where, $\bigtriangledown\,=\,4\,{\partial }^2\,/\,{\partial {\rm E}}\,{\partial {\rm E}^{*}}$. The righthand-side of Eq. (\ref{eq: Poission eq}) vanishes if $G({\rm E})$ satisfies the Cauchy-Riemann conditions and is an analytical function in the complex plane. In other words, we can re-interpret $\rho$ as the measure of non-analyticity in $G({\rm E})$.

As a first step, one must regularise the logarithm in Eq. (\ref{eq: Poission eq}) by introducing a positive infinitesimal $\epsilon$. This ensures that the matrix whose determinant we seek is positive definite; consequently one may represent the determinant \cite{Efetov:1983} as an integral over complex variables,

\begin{eqnarray}
\label{eq: avg potential with integrals}
\nonumber
\langle\,\phi\,\rangle &=&\frac{1}{N}\Bigg\langle\,\ln\,\int \prod_{k=1}^N\left[\frac{{\rm d}^2\,z_k}{2\,\pi}\right]\,\exp\Big\{ -\epsilon\,(z^{*},\,z)\\
&-& (z^{*},\,M\,z)\,\Big\}\Bigg\rangle\,,
\end{eqnarray}
where $(z^{*},\,z)\,=\,\sum_i\,z^{*}_i\,z_i$ and $M\,=\,(\boldsymbol{A}^T\,-\,{\rm E}^{*})\,(\boldsymbol{A}\,-\,{\rm E})$.

The next step is to perform the average $\langle (\ldots) \rangle$ via the replica trick \cite{Edwards:1976}, $\ln\,x\,=\,\lim_{n\,\to\,0}\,(x^n\,-\,1)\,/\,n$, the result of which may be solved in the limit $N\,\to\,\infty$ via the saddle point technique. However, as noted in \cite{Haake:1992}, the integrand that one evaluates has its extremum for order-parameters that do not depend on the replica index. As a result, $\langle\,\phi\,\rangle$ may be calculated formally by setting $n\,=\,1$.

The average is facilitated by linearising terms quadratic in $J$ via a complex Hubbard-Stratonovich transformation,
\begin{eqnarray}
\label{eq: potential with hubbard-strat}
\nonumber e^{N\,\langle\,\phi\,\rangle} &=& \Bigg\langle\,\int\,\prod_{k=1}^{N}\left[\frac{{\rm d}^2\,z_k}{2\,\pi}\,\frac{{\rm d}^2\,y_k}{2\,\pi}\right]\,\exp\Big\{ -\,\epsilon\,(z^{*},\,z) \\
&-& (y^{*},\,y)\Big\} \,\times\,D\,\Bigg\rangle\,,
\end{eqnarray}
where the disorder terms are restricted in
\begin{equation}
\label{eq: disorder}
D\,=\,\exp\left\{ {\rm i}\,(z^{*},\,(\boldsymbol{A}^T\,-\,{\rm E}^{*})\,y)\,+\,{\rm i}\,(y^{*},\,(\boldsymbol{A}\,-\,{\rm E})\,z)\right\}\,,
\end{equation}
and
\begin{equation}
 \label{eq: matrix elements}
a_{ij}\,=\,c_{ij}\frac{J}{\sqrt{c}}\underbrace{\left(x_{ij}\,-\,\frac{1}{|{\cal N}_i|}\sum_k x_{ik}\right)}_{u_{ij}}\,,
\end{equation}

for off-diagonal elements, while $a_{ii}\,=\,-1$. The elements $u_{ij}$ are a linear combination of Gaussian random variables $x_{ij}$ and we verify that up to leading ${\cal O}(1)$,
\begin{equation}
\label{eq: h avg}
\langle u_{ij} \rangle\,=\,0\,,\qquad \langle u_{ij}\,u_{kl} \rangle\,=\,\delta_{i,k}\,\delta_{j,l}\,+\,\Gamma\,\delta_{i,l}\,\delta_{j,k}\,,
\end{equation}
We perform the average of the $c_{ij}$ and $u_{ij}$ in the usual manner; for details refer to \cite{anand:016111}.

We note that in Eq. (\ref{eq: potential with hubbard-strat}) if we rotate all $z_i$ and $y_i$, i.e., multiply them with $\Lambda\,=\,e^{{\rm i}\,\theta}$ that has unit modulus, the integral remains unchanged. More precisely, the terms in the exponential in Eq. (\ref{eq: potential with hubbard-strat}) are invariant under this transformation. In our case, we must take terms of the form $N^{-1}\sum_i (y_i^\star)^2$, $N^{-1}\sum_i (y_i)^2$, $N^{-1}\sum_i (z_i^\star)^2$, $N^{-1}\sum_i (y_i^\star)^2$, $N^{-1}\sum_i z_i^\star\,y_i^\star$ and $N^{-1}\sum_i z_i\,y_i$ all to be equal to zero. We consequently introduce order-parameters
\begin{eqnarray}
\label{eq: order-params 1}
u &=& \frac{1}{N} (z^{*},\,z)\,,\quad v\,=\,\frac{1}{N}(y^{*},\,y)\,,\\
\label{eq: order-params 2}
w &=& \frac{1}{N} (z^{*},\,y)\,,\quad w^{*}\,=\,\frac{1}{N}(z,\,y^{*})\,.
\end{eqnarray}
Consequently, Eq. (\ref{eq: potential with hubbard-strat}) reduces to
\begin{eqnarray}
\label{eq: potential after disorder avg}
\nonumber e^{N\,\langle\,\phi\,\rangle} &=& \int\,\prod_{k=1}^{N}\left[\frac{{\rm d}^2\,z_k}{2\,\pi}\,\frac{{\rm d}^2\,y_k}{2\,\pi}\right]\,\exp\Bigg\{ N \Big[ -\epsilon\,u\,-\,v \\
\nonumber &-&  J^2\,u\,v\,-\,J^2\,\frac{\Gamma}{2}(w^2\,+\,(w^{*})^2)\\
&-&{\rm i}({\rm E}^{*}\,+\,1)\,w\,-\,{\rm i}({\rm E}\,+\,1)\,w^{*} \Big] \Bigg\}\,.
\end{eqnarray}

We introduce the order-parameter definitions Eqs. (\ref{eq: order-params 1}) - (\ref{eq: order-params 2}) into Eq. (\ref{eq: potential after disorder avg}) via Dirac $\delta$ functions. This allows us to perform the integrals over $z_k$ and $y_k$ using Gaussian identities, which yields
\begin{equation}
\label{eq: before saddle pt}
e^{N\,\langle\,\phi\,\rangle}\,=\,\int {\cal D}(\ldots) \exp\{N[\Xi_1\,+\,\Xi_2\,+\,\Xi_3]\}\,,
\end{equation}
where,
\begin{eqnarray}
\nonumber \Xi_1 &=& -\epsilon\,u\,-\,v\,-\,J^2\,u\,v\,-\,J^2\,\frac{\Gamma}{2}[ w^2\,+\,(w^{*})^2]\\
&-& {\rm i}[({\rm E}^{*}\,+\,1)\,w\,+\,({\rm E}\,+\,1)\,w^{*}]\,,\\
\Xi_2 &=& {\rm i}\widehat{u}\,u\,+\,{\rm i}\widehat{v}\,v\,+\,{\rm i}\widehat{w}\,w^{*}\,+\,{\rm i}\widehat{w^{*}}\,w\,,\\
\Xi_3 &=& -\ln [ {\rm i}\widehat{u}\, {\rm i}\widehat{v}\,-\,{\rm i}\widehat{w}\,{\rm i}\widehat{w^{*}} ]\,,
\end{eqnarray}
and ${\cal D}(\ldots)$ denotes the integral over variables the order parameters and their ``hatted" conjugate variables, which are a consequence of a Fourier representation of the Dirac $\delta$ functions.

The average potential, $\langle\,\phi\,\rangle$ is related to the saddle-point value of the function $\Psi\,=\,\Xi_1\,+\,\Xi_2\,+\,\Xi_3$. We proceed by eliminating the conjugate variable by requiring stationarity. Next, we introduce $r\,=\,u\,v\,\geq\,0$ and eliminate $u$ using the stationarity condition ${\rm d}\,\Psi/{\rm d}\,u\,=\,0$.
\begin{eqnarray}
\label{eq: psi no conjugate}
\nonumber \Psi &=& -2\,\sqrt{r\,\epsilon}\,-\,J^2\,r\,-\,J^2\,\frac{\Gamma}{2}[\,(w^{*})^2\,+w^2\,]\\
\nonumber &-& {\rm i}[({\rm E}^{*}\,+\,1)\,w\,+\,({\rm E}\,+\,1)\,w^{*}]\\
&+&2\,+\,\ln[ r\,-\,w\,w^{*} ]\,.
\end{eqnarray}

In the limit $\epsilon\,\to\,0^{+}$, we may distinguish between two possibilities; $\Psi$ may take a maximum at $r\,=\,0$ or there may be an extremum for some $r\,>\,0$. Considering first $r\,=\,0$, Eq. (\ref{eq: psi no conjugate}) resolves to an analytical function in the domain of ${\rm E}$; hence, $G({\rm E})\,=\,0$ and $\rho\,=\,0$.

In the case one obtains an extremum for $r\,>\,0$, with ${\rm E}\,=\,x\,+\,{\rm i}\,y$, we obtain saddle point equations
\begin{eqnarray}
\label{eq: saddle point r}
r &=& \frac{1 \,+\, w\,w^{*}}{J^2}\,,\\
w\,+\,w^{*} &=& \frac{{\rm i}\,2\,(x\,-\,1)}{J^2(1\,+\,\Gamma)}\,,\\
w\,-\,w^{*} &=& -\frac{2\,y}{J^2(1\,-\,\Gamma)}\,.
\end{eqnarray}

Evaluating $\langle \phi \rangle$ at this saddle point and employing Eq. (\ref{eq: Poission eq}) we obtain the average density, $\rho\,=\,1 / \pi\,(1\,-\,\Gamma^2)$, which is valid in the region $r\,>\,0$, i.e.,
\begin{equation}
\label{eq: region of non-analyticity in G}
\frac{(x\,+\,1)^2}{a^2}\,+\,\frac{y^2}{b^2}\,<\,J^2\,,
\end{equation}
where $a=\,1\,+\,\Gamma$ and $b\,=\,1\,-\,\Gamma$.

\bibliographystyle{unsrt}
\bibliography{AnandGalla_Sta_dyn_mat_flo_rnd_net}

\end{document}